\documentstyle[12pt]{article}
\date{}

\title{ EFFECT OF INITIAL PARTICLE INSTABILITY IN MUON COLLISIONS}

\author{I.F.~Ginzburg\\
Institute of Mathematics, Novosibirsk, 630090, Russia\thanks{
Work is supported by grants of International Science Foundation
and Russian Ministry of Sciences (RPL300) and European
association INTAS (INTAS -- 93 -- 1180).}\\ e-mail:
ginzburg@math.nsk.su}
\begin{document}
\maketitle

Preprint DESY--95--168

\begin{abstract}
We study $\mu^+\mu^-$ collision with production of some state $F$
(for example, $F= W^+$) and $(e\bar\nu)$ system with the
effective mass, which is less than muon mass. In this case the
momentum $k$, transferred from $\mu$ to $e\bar\nu$ system
(virtual neutrino momentum), can be time--like. The path of
integration over $k^2$ goes through the pole at $k^2=0$. It gives
divergent cross section.

This divergence disappears if the finite width $\Gamma$ of
initial $\mu$ is taken into account. It gives the factor
$\propto\Gamma^{-1}$ in cross section. The compensating factor
$\Gamma$ appears from $\mu\to e\bar\nu \nu$ subprocess. The
resulting cross section is finite but high enough.
\end{abstract}

The modern discussions about muon colliders (see e.g. \cite
{Palm}) provide new place for studying of effects related to
particle instability.

For the definiteness, we consider the process (momenta of
particles are shown in brackets):
\begin{equation}
\mu^-(p_1)\mu^+(p_2)\to e(q_1)\bar{\nu}(q_2) W^+(p_3).\label{process}
\end{equation}
We use the following notations:\ $M$ is the $W$ boson mass, $m$
is the muon mass, $s\equiv 4E^2 =(p_1+p_2)^2$; $\;x=M^2/s$;
$\;q=q_1+q_2,
\quad k=p_1-q\equiv p_3-p_2$; \ we neglect electron mass.

1. We study the specific kinematical region where the effective
mass of $e\bar{\nu}$ system is less than muon mass, $q^2<m^2$. In
this case the transferred momentum k can be time--like, its
maximal value is positive:
\begin{equation}
k^2< t_{max} =\frac{x}{1-x}\left[m^2 (1-x)-q^2\right] >0.
\label{bound}
\end{equation}
With the growth of total transverse momentum of produced system,
this transferred momentum becomes space--like, $k^2<0$. Therefore
the integration over this transverse momentum (at fixed $q$) goes
through the point $k^2=0$.

The main contribution to the cross section in this region is
given by the diagram with neutrino exchange in t--channel, this
(virtual) neutrino momentum is $k$. This contribution gives a
factor $(k^2)^{-2}$ in the matrix element squared. The standard
integration over $k^2$ results in divergent cross section in this
case\footnote{ I was informed about this paradox by F.~Boudjema.}!

This paradox originates from the instability of muon, decaying
into $e\bar{\nu}\nu$ system. The point $k^2=0, \;q^2 <m^2 $ is
within the physical region for this decay. (The obtained result
is conserved if neutrino have some small mass.  Since muon decays
into $e\nu\bar{\nu}$, the integration over $k^2$ goes through the
pole of neutrino propagator independent of value of neutrino
mass.)

2. This divergence is killed, if one takes into account the fact
that the wave function differs from the standard plane wave due
to muon instability. In muon rest frame:
\begin{equation}
e^{-imt/\hbar}\Rightarrow e^{-i(m-i\Gamma/2)t/\hbar}. \label{wf}
\end{equation}

In this frame the 4-momentum of $\mu^-$ is $\tilde{p}_1
=(m-i\Gamma/2,0,0,0)$. To obtain the energy of produced
$e\bar{\nu}$ system $\tilde{q}^0$ in this system, we use simple
kinematical relations $2p_1q=m^2+q^2-k^2,\;\;$ $2p_1q\equiv
2m\tilde{q}^0$. It results in $\tilde{q}^0= (m^2+q^2-k^2)/2m$.

New value of $k^2$ is obtained from relation $k^2_{new}\equiv
(p_1-q)^2=m^2-im\Gamma +q^2 +i\Gamma\tilde{q}^0$. Using the above
value of $\tilde{q}^0$, we obtain with the necessary
accuracy\footnote{ This way for derivation of $k^2_{new}$ was
proposed by V.G.~Serbo.}:

\begin{equation}
k^2\Rightarrow k^2-i\gamma;\quad \gamma=\frac{m\Gamma (m^2-q^2)}
{2m^2}.\label{change}
\end{equation}

Now we calculate the cross section of the process in the standard
way. Calculations are simplified, since we neglect small
quantities $\sim m^2/M^2$, $\Gamma/m$ in the result.  Finally we
get:
\begin{eqnarray}
d\sigma&=&\frac{|{\cal M}|^2 dk^2 dq^2}{4(4\pi)^3 s(s-4m^2)}
\frac{d\Omega^*_q}{4\pi}\frac{d\varphi}{2\pi};\nonumber\\
|{\cal M}|^2&=&\frac{(4\pi\alpha)^3}{M^2(\sin^2\Theta_W)^3}
\frac{(2q_1 p_1)(2kq_2)}{k^4+\gamma^2}. \label{dsigma}
\end{eqnarray}
Here $d\Omega^*_q$ is the solid angle element in the center of
mass of produced $e\bar{\nu}$ system.

The subsequent angular integration is trivial. The integration
over $k^2$ gives $\pi/\gamma$ for any $q^2$, since the bounds of
the integration region are much higher than $\gamma$. The
integration over $q^2$ covers the region $q^2< m^2(1-x) $, where
exchanged neutrino can be on mass shell (\ref{bound}).  This
procedure is very close to the calculation of the muon decay
width.  We insert this width instead of the corresponding
combination of factors to the eq. (\ref{dsigma}). It compensates the
factor $\Gamma$ in the denominator, and the final result is:
\begin{eqnarray}
\sigma&=&\frac{\pi^2\alpha}{s\sin^2\Theta_W } f(x)
\equiv 20xf(x)\mbox{ nb};\label{crsec}\\
 f(x)&=&4x(1-x)(2-x).\label{flow}
\end{eqnarray}

This equation gives us the solution of discussed problem. With
above prescription we obtain the finite cross section.

Let us discuss briefly some ideas about the interpretation of
this solution.

First, it seems to be natural to consider this phenomenon as an
effect of a real $\mu$ decay. The obtained quantities correspond
to the integration over whole space--time.  This interpretation
is supported by the fact, that the total number of neutrino is
the same as the number of muons, $\int f(x)dx =1$.

The other idea is to consider this effect as some "induced" decay
of muon in the field of spectator. If this approach is correct,
then these muon colliders can be also used as $\mu\nu$,
$\nu\nu$,... colliders with high enough luminosity.

To finally establish which interpretation is really correct, more
detail calculations taking into account sizes of beam and
interaction region are needed.\\

I am grateful for F.~Boudjema, G.~Jikia, G.~Kotkin, K.~Melnikov
and V.~Serbo for the stimulating discussions, to J.~Bartels,
G.~Belanger, W.~Buchmuller, A.~Djouadi, V.~Fadin, V.~Ilyin,
J.~Kwiecinski, F.~Schrempp, M.~Ryskin and P.~Zerwas for the
discussion of results. Besides, I am thankful to W.~Buchmuller
and P.~Zerwas for the hospitality in DESY during finishing of
this work.

\newpage
{\large\bf Appendix}

If the second interpretation is realized, we have very promising
picture.

First of all, the quantity $f(x)$ (\ref{flow}) can be considered as
the neutrino component of the muon. Indeed, our cross section can
be written through the "number of equivalent neutrino" $F(x)dx$
with the energy in the interval $\{xE; (x+dx)E\}$ as $\int F(x)dx
\cdot\sigma_{\mu\nu\to W}(xs)$. The standard approximation for the
cross section of the subprocess $\mu\nu\to W$ with unpolarized
muons is ($S$ is spin of $W$ boson):
$$
\sigma_{\mu\nu\to W}(\hat{s})= (2S+1)4\pi^2\frac{\Gamma_{W\to\mu\nu}}{M}
\delta (\hat{s}-M^2); \quad \Gamma_{W\to\mu\nu}= \frac{\alpha}{12 \sin^2
\Theta_W}.
$$

With these equations, we obtain from eq. (\ref{crsec})
$F(x)\equiv f(x)$, in other words $f(x)dx$ is the "number of
equivalent neutrino" with the energy in the above interval.
(This result was reproduced by V.G.~Serbo by the method, similar
to that for the derivation of equivalent electron method).

The above calculations determine the "flux of equivalent muonic
neutrino". The same approach give fluxes of electrons and
electronic neutrino in muon.

The calculation repeats the above one completely. The effective
$\mu$ decay vertex is reduced, in fact, (with the accuracy
$m^2/M^2$) to the well known form of four--fermion interaction
which admits Pauli--Fierz transformation with exchange electronic
and muonic neutrino: $$
\left(\hat{\mu}^-\gamma^{\alpha}(1-\gamma^5)
\hat{\bar{\nu}}_{\mu}\right)\left(\hat{ e}^+\gamma^{\alpha}(1-\gamma^5)
\hat{\nu}_e\right)=
\left(\hat{\mu}^-\gamma^{\alpha}(1-\gamma^5)
\hat{\nu}_e\right)\left(\hat{ e}^+\gamma^{\alpha}(1-\gamma^5)
\hat{\bar{\nu}}_{\mu}\right)
$$
Therefore, the flux of electronic neutrino in muon coincides with
that for muonic neutrino (\ref{flow}). The same is true for the
electron flux.

For the given mass of produced system, the energy dependence of
the cross section (\ref{crsec}) is determined by the quantity
$xf(x)$. Its highest value is 0.907 at $x=(9-\sqrt{17})/8 = 0.61
$. This quantity is larger than one half of this highest value
for x within interval $\{0.285;0.88\}$. Therefore, high cross
section for the production of some particle with mass $\tilde{M}$
(W boson or charged Higgs or heavy new W boson, etc.) in the
$\nu\mu$ collision is achieved at $\sqrt {s}=1.06\div 1.87\tilde
{M} $. In particular, the of cross section shows that $\mu^+\mu^-
$ collider with $\sqrt {s}=130 $ GeV and luminosity about $10^{31}
\mbox{ cm}^{-2}s^{-1}$ will be the W--boson factory. The number
of produced W's will be here approximately the same as the number
of Z--bosons at LEP.

In this case, the $\mu\mu$ colliders will be simultaneously $\mu
\nu_{\mu}$, $\mu\nu_e$, $\mu e$, $e\nu_{\mu}$, $\nu\nu$
colliders with the same luminosity as in the basic $\mu\mu$
collisions (but with nonmonochromatic new beams).


\begin{thebibliography}{99}
\bibitem{Palm} R.B.~Palmer. Beam Dynamics
Newsletter. \# 8 (1995).

\end{thebibliography}
\end{document}